# Report for the Commission of Inquiry Respecting the Muskrat Falls Project

Prof Bent Flyvbjerg*
Dr Alexander Budzier

*Lead author; all opinions expressed in this report are the opinions of the lead author and he accepts responsibility for all errors and omissions.

August 2018

This report was commissioned by the Commission of Inquiry Respecting the Muskrat Falls Project to provide the national and international context in which the Muskrat Falls Project took place.

The Commission asked for the report to cover three specific topics of questions:

1. What is the national and international context of the Muskrat Falls Project with regards to cost overrun and schedule overrun?
   - What are the typical cost and schedule overruns of hydro-electric dam projects?
   - How do hydro-electric dams compare to other capital investment projects?
   - How do Canadian projects compare to other countries?
2. What are the causes and root causes of cost and schedule overruns?
3. What are recommendations, based on international experience and research into capital investment projects, to prevent cost and schedule overruns in hydro-electric dam projects and other capital investment projects?

**Table of Contents**

# 1 Executive summary

For this report we studied 274 hydro-electric dam projects, in order to place the Muskrat Falls Project into the context of projects in Canada and other countries.

Hydro-electric dam projects are high-risk projects, with an average cost overrun of 96% (median 32%) and an average schedule overrun of 42% (median 27%). Cost and schedule overruns of hydro-electric dam projects have remained constant in the last 60 years. The cost and schedule risks of dams is only exceeded by nuclear power projects.

The data show that cost and schedule overruns are pervasive in capital investment projects. Hydro-electric dams are no exception, neither is Canada.

Often cost and schedule overruns are explained by unforeseen conditions and adverse events, e.g. unforeseen geology, project complexity, scope changes, bad weather. However, these are not root causes. The root causes of cost overruns and schedule delays can be found in optimism and political bias in estimates of geology, complexity, scope, weather, etc., which translate into underestimates of cost and schedule, which later turn into cost and schedule overruns.

The data show that conventional cost and schedule estimates are biased, i.e. systematically underestimating cost and schedule risks. The data do not fit the "error" explanation of overrun, and therefore raise doubts that better models and better data, following this explanation, will improve forecasts.

This leaves optimism and political biases as the best explanations of why cost and schedule are underestimated. Optimism bias and political bias are both deception, but where the latter is deliberate, the former is not. Optimism bias is self-deception.

Project funders, owner/operators, sponsors, project managers, i.e. key decision makers, would be well advised to take the following steps to debias their project plans and proposals:
- Improve project *viability and risk assessments* by taking an outside view of the project, disclose the full distributional information of forecasts, and use Reference Class Forecasting to produce more accurate estimates that bypass optimism and political biases.
- Enhance project *oversight* by making de-biasing of projects part of the stage gate process, conduct independent audits and peer reviews of projects.
- Introduce better *accountability* of planners and forecasters, including aligning positive and negative incentives to produce accurate forecasts and hold project decision makers accountable for project planning and delivery.
- Enhance the *transparency* of project performance. Measuring and reporting project performance against multiple and clearly defined baselines is necessary to hold forecasters accountable for their forecasts, hold decision makers accountable for the quality of their decisions, hold project teams accountable for their project execution and contractors accountable for their contracts. In reporting, special emphasis must be placed on detecting and acting upon early-warning signs so possible damage to the project can be identified and prevented. Better transparency is also needed with regards to unit cost and productivity of projects to ensure value for money.

- The data show that major projects perform best when they are fast and modular and perform worst when they are slow and bespoke. Projects need to scale smartly, i.e. they need to be designed for economies of scale and learning, with as high an element of modularity and speed as possible.

- Finally, maturity of leadership in capital investment projects is often perceived to be lacking. Investment in the development of project leaders, sponsors and stakeholders is necessary to increase the likelihood of project success.

# 2 Cost and schedule overruns

In recent years, hydro-electric dam projects have again figured prominently in energy policies and development agendas. As critics note, hydro-electric dams are in many instances a high-risk strategy.

This report will first address this critique and analyze the cost and schedule overruns of completed hydro-electric dam projects. This analysis uses past data on cost and schedule overruns as the best available predictor of cost and schedule risk of hydro-electric dams.

Lastly, this section compares the cost and schedule overruns of hydro-electric dam projects with other project types; and Canadian projects with projects in other countries.

## 2.1 Cost and schedule overruns of hydro-electric dam projects

Our previous research (Ansar et al. 2014) was based on 245 dams, including 186 hydro-electric dam projects. For this report, we enlarged and updated the sample from 186 to a total of 274 hydro-electric dam projects[1].

Cost overrun is calculated as actual divided by estimated cost. Costs are measured in real terms, i.e. inflation is removed. The estimated cost includes all cost for the build phase of the project estimated at the decision to build, i.e. the full business case. Actual cost include all build cost, but not operating cost, measured at the commencement of revenue operations of the project.

Schedule overrun is calculated as the actual divided by estimated duration of the project from the date of the decision to build to the commencement of revenue operations.

*Table 1 Cost and schedule overruns of hydro-electric dam projects*

|  | Average | Median | Range | Frequency of overrun | Sample size (n) |
|---|---|---|---|---|---|
| *Cost overrun* | +96% | +32% | -47% to +5142% | 77% | 269 |
| *Schedule overrun* | +42% | +27% | -29% to +402% | 80% | 249 |

The data in Table 1 show that cost overrun is more likely than not. Nearly 8 out of 10 past projects incurred a cost overrun.

---

[1] The projects in the analysis comprise the full scope required for the operation of a hydro-electric dam, i.e. civil engineering works for the dam structures, electrical and mechanical installations. In most dams studied the scope also included changes to catchment areas, transmission lines etc. However, as the comparison with transmission projects in Section 2.2.2 below shows, the main source of overrun is the dam itself.

The data also show that on average dams nearly double their budget. The high average is influenced by the presence of outliers in the data. The largest overrun measured was 5,142% (Visegrad Hydroelectric Project 1980-1995).

Outliers are projects with very high cost and/or schedule overruns. These projects are also sometimes called "Black Swans", a popular term for extreme events with massively negative outcomes (Taleb, 2010). In statistical terms, Black Swans are outliers. Outliers are commonly defined to be 1.5 inter-quartile ranges (the difference between the top and bottom quartile) away from the top quartile (Tukey 1977). Defined in this manner, in the data of hydro-electric dam projects outliers are projects with cost overruns ≥ +207% and schedule overruns ≥ +127%. 10% of the observations in the data are classified as cost outliers defined in this manner, 6% of observations are classified as schedule outliers[2].

A common misconception is that Black Swans are freak occurrences to be excluded from risk analyses. However, managers should not ignore Black Swans, because Black-Swan projects are generally not caused by catastrophic risks materializing (e.g. disease outbreaks, terrorism) but are typically the result of multiple adverse events occurring simultaneously. Thus, while they cannot be predicted managers can learn from them to reduce their projects' exposure to Black Swans.

The median cost overrun in the data is 32%. Half of the hydro-electric dam projects have exceeded their cost estimate at the decision to build by more than 32%. The median also represents the typical hydro-electric dam project - typically one third of the estimated cost had to be added between the decision to build and the commencement of operations.

With regards to schedule overrun, the data show that schedule overrun is more likely than not. 8 out of 10 past hydro-electric dams were delayed. Half the dams had a schedule overrun of more than 27%. Based on our data, the average schedule overrun to expect for a hydro-electric dam is 42%, the typical schedule overrun (median) is 27%.

Further, the data show that the mean actual duration of hydro-electric dams is 100 months (approximately 8.3 years) and the median actual duration is 84 months (7 years), measured from the date of decision to build to start of commercial operations. The average delay is 27 months.

Figure 1 shows that historically the average cost and schedule overruns have remained constant. The concerns about the high cost and schedule risk of hydro-electric dam projects are as valid today as they were 60 years ago.

---

[2] 2% of observations were both cost and schedule outliers. These are included in the 10% cost outliers and 6% schedule outliers. This shows that it is more likely that hydro-electric dams had either a large cost overrun or a large delay than having both.

*Figure 1 Historic moving average of cost and schedule overruns in hydro-electric dam projects (logarithmic y-axis to account for outliers, 95% confidence interval of the moving average shown)*

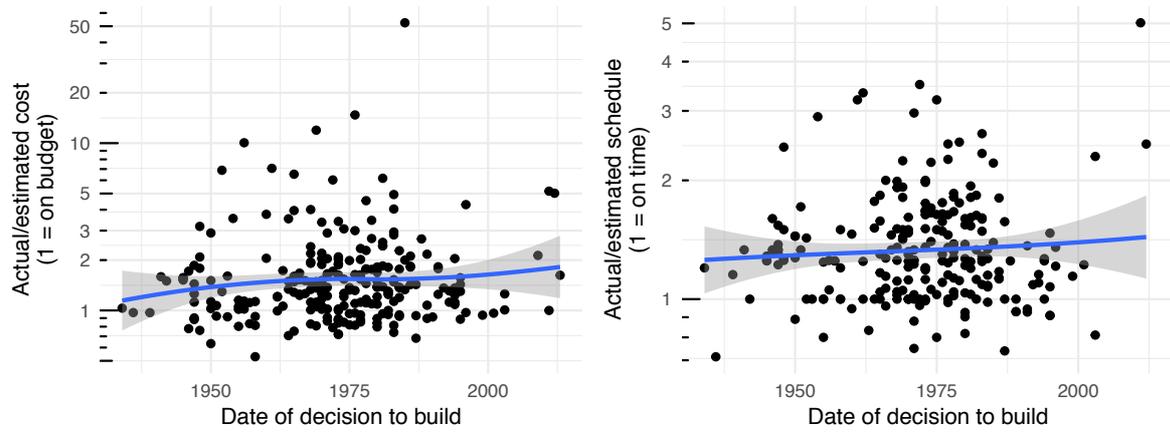

## 2.2 Comparison of hydro-electric dam projects with other capital investment projects

This section of the report compares hydro-electric dam projects to other capital investment projects in transport, energy, and resource extraction (mining and oil & gas). The analysis considers whether hydro-electric dam projects are a special type of project, with regards to cost and schedule overruns.

### 2.2.1 Comparison with transport infrastructure projects

Table 2 shows that the average cost overrun in hydro-electric dams (96%) is statistically significantly greater than the cost overruns in roads and bridges (24% and 32%). Hydro-electric dams have similar cost overrun, i.e. statistically not significantly different, to tunnel (38%) and rail (41%) projects.

The frequency of cost overrun in hydro-electric dam projects is similar to the frequency of cost overrun in transport, where 7-8 out 10 projects have experienced cost overrun.

The average schedule overrun in hydro-electric dams is 42%. In road and tunnel projects the average schedule overrun is statistically significantly lower (20% and 22%). The schedule risk of hydro-electric dam projects is similar to that of bridge (23%) and rail (48%) projects, where the difference is not statistically significant.

*Table 2 Hydro-electric dam projects compared to transport infrastructure projects*

| | Cost overrun (mean) | Frequency of cost overrun | Schedule overrun (mean) | Frequency of schedule overrun | Sample size (n) |
|---|---|---|---|---|---|
| *Hydro-electric dams* | +96% | 77% | +42% | 80% | 274 |
| *Roads* | +24%*** | 72% | +20%*** | 71% | 963 |
| *Bridges* | +32%* | 71% | +23% | 74% | 51 |
| *Tunnels* | +38% | 73% | +22%** | 50% | 56 |
| *Rail* | +41% | 80% | +48% | 80% | 308 |

*** p < 0.001; ** p < 0.01; * p < 0.05 (p-values based on the difference between hydro-electric dam projects and other project types using two-sample Wilcoxon tests)

### 2.2.2 Comparison with energy projects

Table 3 compares hydro-electric dam projects to other energy projects. Dams have the second highest average cost overrun (average 96%). The average cost overrun of dams is statistically significantly higher than those of renewable projects (1% and 13%), transmission projects (8%) and statistically significantly higher than conventional power plants using coal, gas, oil or diesel as power source (31%).

The average cost overrun of hydro-electric dams are only exceeded by nuclear power projects, which had an average cost overrun of 122%. Again, this difference is statistically significant.

Similarly, the average schedule overrun of hydro-electric dams (44%) statistically significantly exceeds the average schedule overruns of renewables projects (0% and 22%) and transmission projects (8%). The average schedule overrun of dams is statistically significantly smaller than that of nuclear projects (65%). The average schedule overrun of hydro-electric dams is similar to the average schedule overrun of thermal power generation projects (36%); the difference is not statistically significant.

*Table 3 Hydro-electric dam projects compared to energy projects*

|  | Cost overrun (mean) | Frequency of cost overrun | Schedule overrun (mean) | Frequency of schedule overrun | Sample size (n) |
|---|---|---|---|---|---|
| Hydro-electric dams | +96% | 77% | +44% | 80% | 274 |
| Wind power | +13%*** | 64% | +22%* | 64% | 53 |
| Solar power | +1%*** | 41% | -0%*** | 22% | 39 |
| Thermal (oil, gas, diesel, coal) | +31%*** | 59% | +36% | 76% | 124 |
| Transmission | +8%*** | 40% | +8%*** | 12% | 50 |
| Nuclear | +122%*** | 97% | +65%*** | 93% | 191 |

*** $p < 0.001$; ** $p < 0.01$; * $p < 0.05$ (p-values based on the difference between hydro-electric dam projects and other project types using two-sample Wilcoxon tests)

## 2.2.3 Comparison with oil, gas and mining projects

For oil, gas and mining projects, i.e. resource extraction, the sample did not include sufficient data points to analyze schedule overrun. The average cost overrun in hydro-electric dam projects (96%) is statistically significantly higher than the average overrun of 17% in resource extraction projects.

*Table 4 Hydro-electric dam projects compared to oil, gas and mining projects*

|  | Cost overrun (mean) | Frequency of cost overrun | Sample size (N) |
|---|---|---|---|
| Hydro-electric dams | +96% | 77% | 274 |
| Mining, oil & gas | +17%*** | 60% | 531 |

*** $p < 0.001$; ** $p < 0.01$; * $p < 0.05$ (p-values based on the difference between hydro-electric dam projects and mining, oil & gas projects using two-sample Wilcoxon tests)

## 2.3 Comparison of Canadian projects with projects in other countries

The comparison of Canadian hydro-electric dam projects with projects constructed in other countries (Table 5) shows that the average cost overrun is lower in Canada (41%) than it is in other countries (99%). Although the difference in the average cost overruns is large, variations in the data mean that it is *not* statistically significant.

Schedule overrun is also lower, with Canadian hydro-electric dam projects being delayed on average by 13% and 43% elsewhere. This difference is statistically significant.

Table 5 shows that Canadian hydro-electric dam projects had a lower schedule, but not cost, overrun compared to the rest of the world.

*Table 5 Comparison of hydro-electric dam projects in Canada with other countries*

|  | Cost overrun (mean) | Frequency of cost overrun | Schedule overrun (mean) | Frequency of schedule overrun | Sample size (n) |
|---|---|---|---|---|---|
| Canada | +41% | 50% | +13%* | 50% | 19 |
| Rest of the world | +99% | 78% | +43%* | 81% | 254 |

*** $p < 0.001$; ** $p < 0.01$; * $p < 0.05$ (p-values based on the difference between hydro-electric dam projects in Canada and in other countries types using two-sample Wilcoxon tests)

Table 6 compares Canadian transport, energy (excluding hydro-electric dams) and resource extraction (mining, oil & gas) with the same type of project in other countries.

In transport and non-hydro energy projects the projects in Canada had a similar average cost overrun to the overrun experienced elsewhere. While the Canadian average cost overrun is slightly lower in both categories the difference is *not* statistically significant.

In mining, oil and gas projects Canadian projects have statistically significantly lower cost overruns (Canada 13%, rest of the world 44%).

When considering schedule overrun, Canadian transport projects have a statistically significantly lower overrun. Canadian energy projects are similar to those in other countries, with regards to schedule overrun (no statistically significant difference). The analysis did not have sufficient data to compare schedule overrun for mining, oil & gas projects.

*Table 6 Comparison of Canadian projects with projects in other countries (transport, energy, mining, oil & gas)*

| Project type | Location | Cost overrun (mean) | Frequency of cost overrun | Schedule overrun (mean) | Frequency of schedule overrun | Sample size (n) |
|---|---|---|---|---|---|---|
| *Transport* | *Canada* | +20% | 60% | +4%** | 42% | 21 |
| | *Rest of world* | +29% | 74% | +42%** | 77% | 1365 |
| *Energy (excluding hydro-electric)* | *Canada* | +74% | 83% | +46% | 57% | 24 |
| | *Rest of world* | +79% | 76% | +41% | 74% | 633 |
| *Mining, oil and gas* | *Canada* | +13%*** | 56% | +16% | 81% | 458 |
| | *Rest of world* | +44%*** | 85% | NA | NA | 73 |

*** $p < 0.001$; ** $p < 0.01$; * $p < 0.05$ (p-values based on the difference between Canadian projects and projects in other countries of the same type using two-sample Wilcoxon tests; NA = not available)

# 2.4 Summary of the findings

The key findings of the analysis were:

- Average cost overrun of hydro-electric dam projects is 96% (median 32%)
- Average schedule overrun of hydro-electric dam projects is 42% (median 27%)
- Cost and schedule overruns of hydro-electric dam projects have remained constant in the last 60 years
- Hydro-electric dam projects have statistically significantly higher cost overruns than road and bridge projects in transport; wind, solar and thermal power plant projects in energy; and mining, oil & gas projects.
- Cost overrun of hydro-electric dam projects are similar, i.e. not statistically significantly different, to rail and tunnel projects.
- Hydro-electric dam projects only have statistically significantly lower cost overruns than nuclear power plants.
- Hydro-electric dam projects have statistically significantly higher schedule overrun compared with road and tunnel projects; and wind and solar power projects.
- Hydro-electric dam projects have a similar schedule overrun as bridges and rail; thermal power plants (i.e. they are not statistically significantly different).
- The only project type with statistically significantly greater schedule overrun is nuclear power.
- With regards to cost overrun, Canadian hydro-electric dam, transport, energy projects are similar (i.e. not statistically significantly different) to projects in other countries

- Canadian cost overruns are statistically significantly lower in mining, oil & gas projects compared to similar projects in other countries.
- With regards to schedule overrun, Canadian hydro-electric dam and transport projects have statistically significantly lower overruns as projects in other countries.
- Schedule overruns are similar in Canadian energy projects (excluding hydro-electric dams).

The data show that cost and schedule overruns are pervasive in capital investment projects. Hydro-electric dams are no exception, neither is Canada.

The data show that hydro-electric dam projects are high risk; only nuclear power plants have had greater cost and schedule overruns.

Next, the report is going to analyze the causes and root causes of cost and schedule overruns, before turning to recommendations of how this situation can be improved.

# 3 Causes of cost and schedule overruns

This section analyzes the causes and root causes of cost and schedule overruns. First, this section is going to look at the official explanations of cost and schedule overrun that were given by the Niagara Tunnel Project. This section is then exploring the underlying root causes in the Niagara Tunnel Project and other projects.

## 3.1 Niagara Tunnel Project Case

In 2004, the Niagara Tunnel Project was sanctioned by Ontario Power Generation (OPG). OPG estimated the 10.2 km tunnel to cost CAD 985.2 millions and to complete in the fall of 2009.

In March 2013, OPG announced completion of the tunnel and declared the project in service. The actual outlay was CAD 1.5 billion (62% increase). Completion was delayed by 42 months against the original business case (OPG 2013).

The original budget was informed by a quantified risk analysis. For the tunneling contract a cost contingency of CAD 96 million and a schedule contingency of 36 weeks were allocated to provide 90% certainty that the targets would be met (P90). The overall project cost contingency was set at CAD 112 million, included in the CAD 985.2 million budget.

The project was delayed on several occasions. OPG cited as reasons for the delay slower than expected progress of the tunnel boring machine (TBM) – 6.06 m/day instead of 14.55 m/day – due to the rock conditions encountered (OPG 2013, p. 70). When the tunneling contract was renegotiated in 2009, OPG updated the cost estimate to CAD 1.6 billion and explained:

"Some of the primary drivers cited for the schedule [and cost] variances are:
- Slower than planned TBM progress due to worse than expected conditions in the Queenston shale once the tunnel passed the St. Davids Gorge.
- Expectation of continuing challenges as the tunnel ascends to higher rock strata and undertakes more mixed-face mining. […]
- Restoring the tunnel to a circular profile ("profile restoration") is an additional task that was not included in the original schedule. […]
- Additional time to allow for removal of tunneling equipment before removal of the cofferdam at the intake structure." (OPG 2013, pp. 112-113)

OPG's explanation of the cost increase and delay of the Niagara Tunnel Project is typical of the explanation provided by projects once they experience cost and schedule overruns.

## 3.2 Common causes of cost and schedule overruns

Similar to the explanations given by OPG for the Niagara Tunnel Project, funders, owner-operators and builders of projects tend to explain cost and schedule overruns in major projects as a result of unforeseen ground conditions, project complexity, scope and design changes, weather, delays in site access and possession, delays in obtaining permits etc. (see Cunningham 2017, for a review of studies of causes of cost and schedule overruns).

No doubt, all of these factors at one time or another contribute to cost overrun and schedule delay, but it may be argued that they are not the real, or root, cause. The root cause of overrun is the fact that project planners tend to systematically underestimate or even ignore risks of complexity, scope changes, etc. during project development and decision making.

The root cause of cost overrun and schedule delay is not that unforeseen conditions and adverse events happen to a project. The root cause is found in what a project did or did not do to prepare for unforeseen conditions and adverse events.

## 3.3 Root causes of cost overruns and schedule delays

Most projects change in scope during progress from idea into reality. Changes may be due to uncertainty regarding the level of ambition, the exact corridor, the technical standards, safety, environment, project interfaces, geotechnical conditions, etc. In addition, the prices and quantities of project components are subject to uncertainty.

Hence, some degree of cost and schedule *risk* will always exist. Such risk is however not unknown and should be duly estimated and reflected in the project documentation at any given stage. Hence, cost overruns and schedule delays should be viewed as *underestimation* of cost and schedule risk.

Only identifying the root causes of what causes projects to underestimate cost and schedule risk allows planners and decision makers to address the issue.

At the most basic level, the root causes of cost overrun and schedule delay may be grouped into three categories, each of which will be considered in turn: (1) bad luck or error; (2) optimism bias; and (3) strategic misrepresentation.

### 3.3.1 Error

Bad luck, or the unfortunate resolution of one of the major project uncertainties mentioned above, is the explanation typically given by management for a poor outcome. The problem with such explanations is that they do not hold up in the face of statistical tests.

Explanations that account for overruns in terms of bad luck or error have been able to survive for decades only because data on project performance has generally been of low quality, i.e. data has been disaggregated and inconsistent, because it came from small-N samples that did not allow rigorous statistical analyses. Once higher-quality data was established that could be consistently compared across projects in numbers high enough to establish statistical significance, explanations in terms of bad luck or error collapsed. The very high levels of statistical significance in Table 7 show that such explanations simply do not fit the data.

*Table 7 Tests of the "error" explanation for hydro-electric dams*

|  | Mean | Wilcoxon test, whether the error centers on zero | Frequency of overrun | Binomial test, whether overruns are as frequent as underruns |
|---|---|---|---|---|
| Cost overrun | 96% | p < 0.001 | 77% | p < 0.001 |
| Schedule overrun | 42% | p < 0.001 | 80% | p < 0.001 |

First, if underperformance was truly caused by bad luck and error, we would expect a relatively unbiased distribution of errors in performance around zero. In fact, the data show with very high statistical significance that the distribution does not center on zero and that the forecasting error is biased towards overrun.

Second, if bad luck or error were main explanations of underperformance, we would expect an improvement in performance over time, since in a professional setting errors and their sources would be recognized and addressed through the refinement of data, methods, etc., much like in weather forecasting or medical science.

Substantial resources have in fact been spent over several decades on improving data and methods in major project management, including in cost and schedule forecasting. Still the evidence shows (see Figure 1) that this has not led to improved performance in terms of lower cost overruns and delays.

Bad luck or error, therefore, do not appear to explain the data.

### 3.3.2 Optimism bias

Psychologists tend to explain the underestimation of cost and schedule risks in terms of optimism bias, that is, a cognitive predisposition found with most people to judge future events in a more positive light than is warranted by actual experience. Kahneman and Tversky's (1979a, b) found that human judgment is generally optimistic due to overconfidence and insufficient regard to distributional information about outcomes.

Thus people will underestimate the costs, completion times, and risks of planned actions, whereas they will overestimate the benefits of the same actions. Similarly, the cost and time needed to complete a project will be optimistic, i.e. under estimated. Such errors of judgment are shared by experts and laypeople alike, according to Kahneman and Tversky.

From the point of view of behavioral science, the mechanisms of scope changes, complex interfaces, archaeology, geology, bad weather, business cycles, etc. are not unknown to planners of capital projects, just as it is not unknown to planners that such mechanisms may be mitigated, for instance by Reference Class Forecasting (see below).

However, planners often underestimate these mechanisms and mitigation measures, due to overconfidence bias, the planning fallacy, and strategic misrepresentation. In behavioral terms, scope changes etc. are manifestations of such underestimation on the part of planners, and it is in this sense that bias and underestimation are the root causes of cost overrun. But because scope changes etc. are more visible than the underlying root causes, they are often mistaken for the cause of cost overrun.

In behavioral terms, the causal chain starts with human bias which leads to underestimation of scope during planning which leads to unaccounted for scope changes during delivery which lead to cost overrun. Scope changes are an intermediate stage in this causal chain through which the root causes manifest themselves.

With behavioral science we say to planners, "*Your biggest risk is you.*" It is not scope changes, complexity, etc. in themselves that are the main problem; it is how human beings misconceive and underestimate these phenomena, through overconfidence bias, the planning fallacy, etc. This is a profound and proven insight that behavioral science brings to capital investment planning.

Behavioral science entails a change of perspective: *The problem with cost overrun is not error but bias*, and as long as you try to solve the problem as something it is not (error), you will not solve it. Estimates and decisions need to be de-biased, which is fundamentally different from eliminating error (Kahneman et al. 2011, Flyvbjerg 2008, 2013).

Furthermore, *the problem is not even cost overrun, it is cost underestimation*. Overrun is a consequence of underestimation, with the latter happening upstream from overrun, often years before overruns manifest. Again, if project planners and decision makers try to solve the problem as something it is not (cost and schedule overruns), you will fail. Planners and decision makers need to solve the problem of cost underestimation to solve the problem of cost overrun. Until these basic insights from behavioral science are understood, it is unlikely to get capital investments right, including cost and schedule estimates.

### 3.3.3 Political bias

Economists and political scientists tend to explain underreporting of budget and schedule risks in terms of strategic misrepresentation, or political bias (Wachs 1989, Flyvbjerg 2005). Here, when forecasting the outcomes of projects, forecasters and planners deliberately and strategically overestimate benefits and underestimate cost and schedule in order to increase the likelihood that it is their projects, and not the competition's, that gain approval and funding.

According to this explanation, actors purposely spin scenarios of success and gloss over the potential for failure. This results in managers promoting ventures that are unlikely to come in on budget or on time, or to deliver the promised benefits.

Political bias can be traced to political and organizational pressures, for instance competition for scarce funds or jockeying for position, and to lack of incentive alignment.

The key problem that leads to political bias is a lack of accountability for the parties involved in project development and implementation:

(1) Because of the time frames that apply to major project development and implementation, politicians involved in producing overoptimistic forecasts of project viability in order to have projects approved are often not in office when actual viability can be calculated.

(2) Special interest groups can promote projects at no cost or risk to themselves. Others will be financing the projects, and often taxpayers' money is behind them, including in the form of sovereign guarantees. This encourages rent-seeking behavior for special interest groups.

(3) Contractors, who are an interest group in its own right, are eager to have their proposals accepted during tendering. Contractual penalties for producing over-optimistic tenders are often low compared to the potential profits involved. Therefore, costs and risks are also often underestimated in tenders. The result is that real costs and real risks often do not surface until construction is well under way.

Explanations of cost and schedule overruns in terms of political bias account well for the systematic underestimation of costs and schedule found in the data. A politically biased estimate of costs would be low, resulting in cost overrun, a politically biased estimate of schedule would be short, resulting in delays.

Optimism bias and political bias are both deception, but where the latter is deliberate, the former is not. Optimism bias is self-deception.

## 3.4 Summary of the root causes

Research into the track record of past estimates (e.g. Flyvbjerg et al. 2004, Flyvbjerg 2014, 2016) shows that project cost and schedule estimates are systematically and consistently lower than actual outturn cost and actual schedule.

The data show that conventional, inside-view cost and schedule estimates are biased, i.e. they systematically underestimate cost and schedule risks. The data do not fit the "error" explanation of overrun and raise doubts that better models and better data on their own will improve forecasts.

This leaves optimism and political bias as the best explanations of why cost and schedule are underestimated.

As illustrated schematically in Figure 2, explanations in terms of optimism bias have their relative merit in situations where political and organizational pressures are absent or low, whereas such explanations hold less power in situations where political pressures are high.

*Figure 2 Optimism and Political Bias*

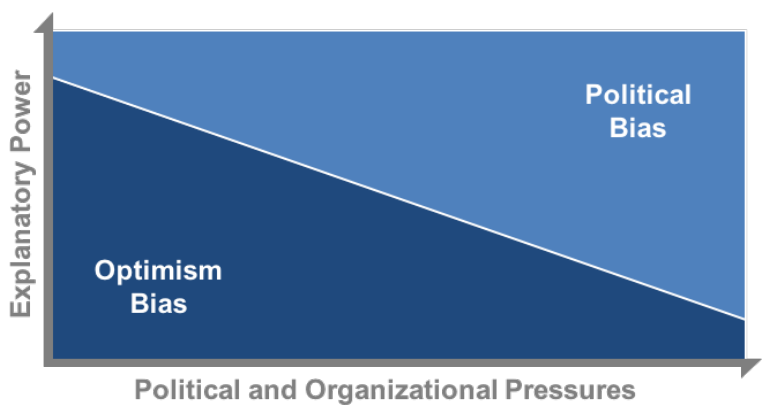

Conversely, explanations in terms of strategic misrepresentation have their relative merit where political and organizational pressures are high, while they become less relevant when such pressures are not present.

Although the two types of explanation are different, the result is the same: inaccurate forecasts and inflated benefit-cost ratios.

Thus, rather than compete, the two types of explanation complement each other: one is strong where the other is weak, and both explanations are necessary to understand the pervasiveness of inaccuracy and risk in project budgeting and scheduling – and how to curb it.

# 4 Recommendations

This section outlines key recommendations on how to de-bias projects based on international experience.

## 4.1 Viability and risk assessments

The research, discussed above, showed that the *causes* of cost overrun and schedule delay can be found within the conventional explanations of why overruns occurred: unforeseen ground conditions, project complexity, bad weather etc.

However, as argued above, the *root cause* of why unforeseen conditions and adverse events turn into overruns can be found in optimistic or political bias in estimates. These underestimations later turn into overruns.

Project funders, owner/operators, sponsors, project managers i.e. key decision makers in projects, should take the following steps to debias their project plans and proposals.

### 4.1.1 Take an outside view

The conventional *"inside view"* of project planning and evaluation results in optimistic estimates and plans. Planners and decision makers with an "inside view" focus on the constituents of the specific planned action rather than on the outcomes of similar actions that have already been completed, i.e. an *"outside view"*.

The outside view pools lessons from past projects. In the basic form, the outside view can be taken by comparing the project at hand to comparable past projects with a view to learn from them.

Projects are typically weak in applying lessons learned from other projects. Research has shown that this is linked to the perceived uniqueness of projects. When project planners perceive their project to be unique they implicitly exclude the experience and knowledge gained from other projects because these are not relevant to their project. In reality, unique projects are rare. Projects are typically specific to a location and a context, but they are rarely unique when looking at global experience and track record.

Thus as a first step, decision makers should challenge and evaluate the quality of estimates and plans by taking the outside view of their project.

## 4.1.2  Probabilistic forecasts of risk

Research has shown that even when project take an outside view, they tend to be biased when presenting projects as single point estimates, i.e. when estimates ignore the full distribution of possible outcomes.

The industry standard of quantitative risk assessments has evolved to present estimates as distributions through Monte Carlo simulations. However, the full distributional information of these quantitative risk assessments is not always shared with decision makers. More importantly, Monte Carlo simulations are not a tool that automatically de-biases risk estimates. Monte Carlo simulations based on optimistic and politically biased inputs create biased forecasts. Garbage in, garbage out, here as elsewhere.

During the front end, when projects are appraised, three key questions are usually considered:
- Is the project economically viable?
- Is the project affordable?
- What project budget and timeline should be set?

The risk appetite of decision makers and hence the total estimate will differ for each of these questions. Sponsors and funders should use probabilistic forecasts instead of single point forecasts to capture this reality.

For example, the question of *economic viability* is relevant to economic appraisals of projects. For this question the *mean* of the quantitative risk assessment is the recommended measure. The mean reflects the expected cost, schedule and benefits of when a project, that is part of a large portfolio of investments, will deliver the outcome intended.

When evaluating project *affordability*, which is a key concern not only in publicly funded projects, decision makers tend to require a higher degree of certainty, i.e. they have a low risk appetite. To evaluate the affordability, decision makers should consider a *downside scenario*, i.e. estimates at a high P-level (P80-P90). In some instances, e.g. the UK's High Speed 2 Project, decision makers have asked for a 95% level of certainty of estimates (P95) to evaluate the affordability and judge whether a project could bankrupt departments or private sector partners.

Lastly, when setting the targets for *budgets and timelines* decision makers need to trade-off between the level of certainty required and the level of challenge and ambition set for suppliers and builders of a project. In practice, a *tiered contingency regime* is becoming the standard approach to achieve this trade-off between control and ambition.

*Figure 3 Tiered Contingency Regime Using a Probabilistic Forecast*

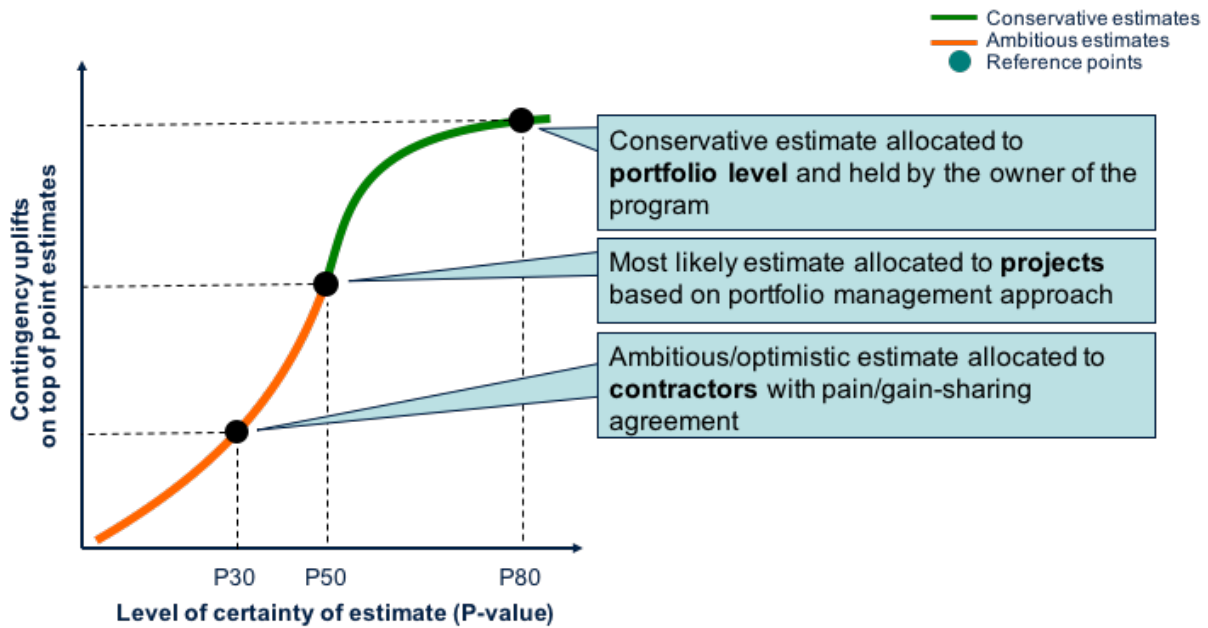

The full distributional information of a forecast could be used to design a tiered contingency regime as shown in Figure 3. For example, a contingency regime could consist of:

-   Contract contingency up to P30: small contingency allocated to key contracts with authority delegated to the contract manager, setting ambitious targets for contractors with downward pressure on costs and demonstrating efficient use of taxpayer money;
-   Project contingency up to P50: additional contingency whose spending authority is delegated to the project manager and which anchors the total cost of the project at the most likely cost estimate;
-   Funder's contingency up to P80: additional contingency whose spending authority is delegated to the project funder or project board, which covers cost above the most likely estimate and includes extreme downside scenarios.

The key advantages of a contingency regime designed in this way are that:

1.  Contractors and contract managers are given an aspirational target. Decision makers are able to set ambitious goals to safeguard value-for-money and incentivize contractors to be cost efficient and innovative;
2.  The project is given a target in line with the likely cost, which follows common planning practice, i.e. uses most likely schedule and cost estimates, and holds project managers to account for their plans; and
3.  The funders of the project reserve a contingency reflecting their level of, typically low, risk appetite.

Each of the three parties should also be given incentives, positive and negative (pain-gain sharing), to achieve their target. For example, UK Department for Transport guidance to local authorities (DfT 2011) states that the department first looks to local authorities to fund any cost increases above their estimate.

Secondly, the department will normally not consider supporting more than 75% of any cost increase. In effect, this sets strong incentives to local authorities with regards to the accuracy of cost risk estimates through establishing an approval process for cost increases and ensuring that local authorities have skin in the game.

In another example, the regime at Heathrow's Terminal 5 set ambitious target costs for contracts. An independent cost auditor verified those target costs and their achievement. Cost savings below a target cost were used to replenish the contingency budget; works above target cost were paid from the contingency. At project completion, contractors received a share of the unspent contingency as a bonus.

A tiered contingency regime like those described above creates transparency about the risks taken on by each party working on a project. These regimes also introduce incentives that motivate each party to deliver according to their estimates and increases the likelihood of delivering project on budget and on time.

## 4.1.3 Data-driven realistic assessments of risk with Reference Class Forecasting

More accurate estimates and thus higher-quality project decisions combine the *"outside view"* and the use of *all the distributional information* that is available. This may be considered the single most important piece of advice regarding how to increase accuracy in forecasting through improved methods, according to Kahneman (2011).

Reference Class Forecasting is a method for systematically taking an outside view on planned actions. Reference class forecasting places particular emphasis on relevant distributional information because such information is most significant to the production of accurate forecasts.

Reference Class Forecasting makes explicit, empirically based adjustments to estimates. In order to be accurate, these adjustments should be based on data from past projects or similar projects elsewhere, and adjusted for the unique characteristics of the project in hand.

Reference Class Forecasting follows three steps:

1. Identify a sample of past, similar projects – typically a minimum of 20-30 projects is enough to get started, but the more projects the better;
2. Establish the risk of the variable in question based on these projects – e.g. identify the cost overruns of these projects; and
3. Adjust the current estimate – through an uplift or by asking whether the project at hand is more or less risky than projects in the reference class, resulting in an adjusted uplift.

It should be noted that any adjustments to the uplift in the final step ought to be based on hard evidence in order to avoid reintroducing optimism back into the estimate.

Because Reference Class Forecasts are based on the actual outcomes of similar past projects, the method estimates not only the known unknowns of a project, i.e. risks identified ex-ante, but also the unkown-unknowns for the project, i.e. risks that have not been identified but may nevertheless impact the project.

### 4.1.4 Example of Reference Class Forecast

For example, a reference class forecast of cost risk of a future hydro-electric project could be based on the data which were analyzed in Section 1.

*Step 1 – Identify a sample of past, similar projects.* The analysis in Section 1 showed, while the average cost overrun was smaller in Canada, the difference was not statistically significant due to the variation in the data. Thus, no convincing statistical evidence exist that any data from other countries should be excluded and therefore all data points should be included to not throw out valuable information.

*Step 2 – Establish the risk of the variable in question.* The variable in question here is cost overrun. The available data are sorted from smallest to largest overrun and the cumulative frequency is calculated. The distribution (Figure 4) shows that cost overrun up to 40% was observed in 52% of projects; and that a cost overrun of up to 100% occurred in approximately 80% of projects.

*Figure 4 Cumulative frequency of cost overrun observed in the data on hydro-electric dam projects (n=274)*

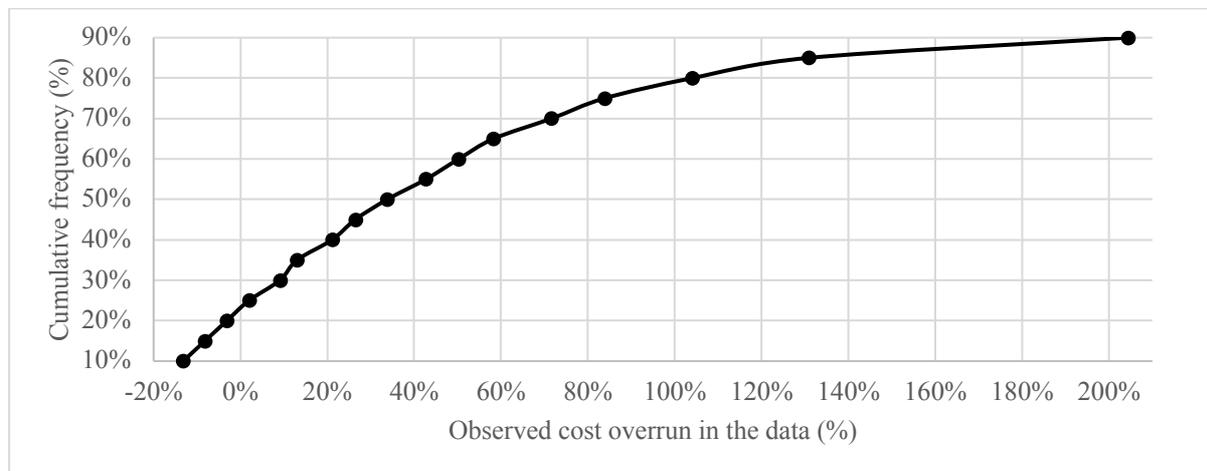

*Step 3 – Adjust the current estimate.* If the project is no more or less risky than similar past projects, the reference class forecast provides the uplift necessary to de-bias the underestimation of cost risk.

To identify the necessary uplift, the data in Figure 4 are re-drawn with both axes swapped. The new x-axis, which was the y-axis in Step 2 and which showed the cumulative of frequency of projects, now has a new meaning: the axis shows the level of certainty required by decision makers for the forecast. The new y-axis, which was the x-axis in Step 2 and which showed the size of cost overrun, is now the required cost uplift, i.e. the relative value of the underestimation of cost risk.

*Figure 5 Cost uplifts to be applied to a hydro-electric project based on the desired level of certainty of decision makers (n=274)*

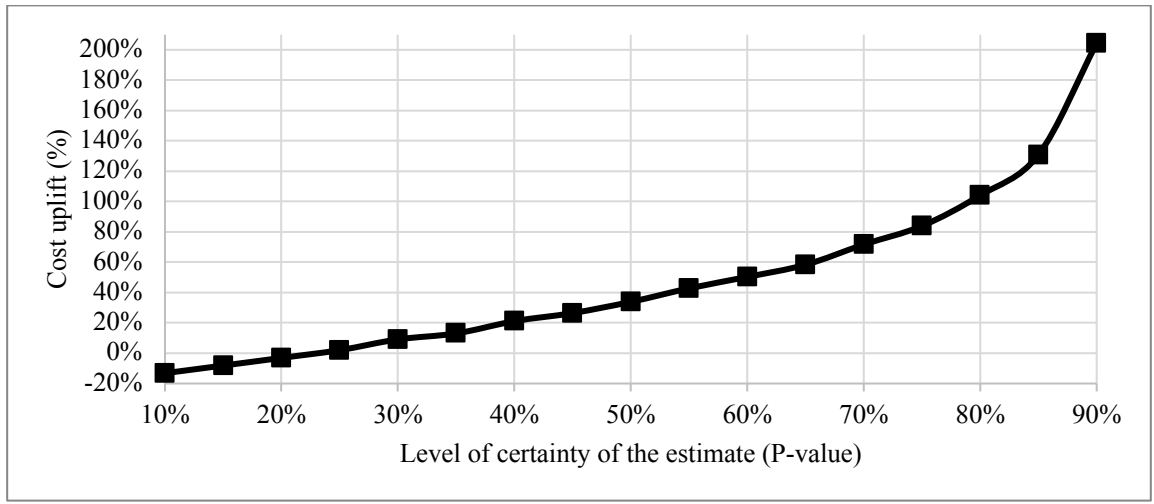

Figure 5 allows decision makers to choose their risk appetite by choosing the level of certainty required for the cost risk estimate. For example, the 50% certain estimate (P50) is 34%. Thus if the forecasted cost are uplifted by 34% the new budget will be met with a probability of 50% and exceeded with a probability of 50%, assuming that the proposed project is no more and no less risky than past, similar projects.

The P50 estimate is often used to forecast projects in a portfolio of projects, because in this manner on average underruns will compensate for overruns and the portfolio will balance overall. However, for big, one-off capital investment projects, decision makers will typically regard a level of 50% certainty to be too low. In this case, decision makers would typically want estimates with a higher level of certainty for staying on budget, often 80% certainty (P80), i.e. estimates with a 20% probability of being exceeded. An 80% certain estimate, Figure 5 shows, requires an uplift of 104%. In this risk averse scenario, decision makers would have to apply a 104% uplift to their project proposal to ensure that the probability of a budget overrun is reduced to 20%. In some cases decision makers have asked for even higher levels of certainty than 80%, for instance 95% (P95) for UK's High Speed 2.

In practice, some decision makers are concerned about large contingencies. They fear what has been called the "red-meat syndrome", i.e. that the mere fact that contingencies are available will incentivize behavior with contractors and others that means the contingencies will be spent. The data for hydro-electric dams and other large projects show clearly that even large contingencies are not excessive but realistic. Instead of avoiding realistic contingencies projects need to put in place incentive schemes (see above), accountability and transparency (see below) to ensure that contingencies are spent only if and when needed, so the "red-meat syndrome" may be avoided. Good project leaders know how to do this.

# 4.2 Oversight

Project oversight and governance are commonly executed through regular project reviews, typically at stage gates, which authorize funding for the next stage. Other governance meetings, e.g. steering committees, are used to review progress and risks, to address arising issues and to make decisions as needed.

With regard to de-biasing project plans and decisions, the key to productive and constructive oversight is to provide critical challenge of project forecasts. The following recommendations focus on enhancements of standard governance structures and processes to de-bias projects.

## 4.2.1 Make de-biasing part of the stage gate/approval process

In 2003 the UK Government introduced the concept of Reference Class Forecasting as part of the HM Treasury Green Book approval process for projects. In the UK the procedure is also known as Optimism Bias Uplifts and typically only applied to cost estimates. Although guidance also recommends the application to schedule and benefits estimates, this is rarely practiced.

The Green Book states: "To redress this [optimistic] tendency, appraisers should make explicit adjustments for this bias. These will take the form of increasing estimates of the costs and decreasing, and delaying the receipt of, estimated benefits. Sensitivity analysis should be used to test assumptions about operating costs and expected benefits" (HM Treasury 2003: 29). And further: "Adjustments should be empirically based, (e.g. using data from past projects or similar projects elsewhere), and adjusted for the unique characteristics of the project in hand. Cross-departmental guidance for generic project categories is available, and should be used in the absence of more specific evidence" (HM Treasury 2003: 29).

Similarly, the Hong Kong Development Bureau started to introduce Reference Class Forecasting for cost estimates in 2014. The UK Government uses broad, generic reference classes available; the Hong Kong Government has built reference classes specific to each individual department.

The key enforcing mechanism, in both the UK and Hong Kong cases, is that projects are forced to compare their inside view with an outside view at key approval stages. In the UK context this is at the approvals for Strategic Outline Business Case, Outline Business Case and Full Business Case. In Hong Kong the approval gates are Upgrade to Category C (inclusion in agency's plan), Upgrade to Category B (completion of feasibility study), and Upgrade to Category A (final decision to build after detailed design and environmental impact and risk assessment).

Three important points need to be considered when integrating Reference Class Forecasting with the stage gate/approval process for projects.

First, the choice of baseline from which overrun is measured is important. To de-bias cost overrun or schedule overrun, the reference class needs to measure the variable of interest against the same baseline for all projects, including the one that is being forecasted.

The same baseline means that, for example, for a cost risk forecast at Outline Business Case approval, the reference class needs to be based on cost overrun data which measures actual cost against estimated cost at Outline Business Case approval. The most common error in Reference Class Forecasts is that data based on contract variations are used for decisions at earlier baselines (e.g. outline or full business case approval stage), leading to significant underestimates of cost and schedule.

Second, during a project's planning process increasing levels of detail become known to planners and decision makers as time passes. This often creates the expectation that risks are reducing. As the data above show, this is not supported by the evidence; sizeable risks remain even in full business case estimates and later.

Thus guidance needs to be given as to how projects combine their inside and outside view risk estimates. The HM Treasury Green Book states: "It is good practice to add a risk premium to provide the full expected value of the Base Case. ... [I]n the early stages of an appraisal, this risk premium may be encompassed by a general uplift to a project's net present value, to offset and adjust for undue optimism. But as appraisal proceeds, more project specific risks will have been identified, thus reducing the need for the more general optimism bias [uplift]" (HM Treasure 2003: 29).

To further clarify this relationship between identified risk and optimism bias uplifts, the UK Government has published a guidance (HM Treasury 2015), which includes Figure 6. In the front-end process a project is expected to identify and plan for specific risks, thus the gap between outside (Reference Class Forecast) and inside view (Quantitative Risk Assessment) should be narrowing. Yet the gap will never fully close because a certain level of unknown risk will always be present in a project's plan.

*Figure 6 Quantitative risk assessment (QRA) and Reference Class Forecasting (RCF) over the lifecycle of a project (Source: HM Treasury Infrastructure Routemap 2015).*

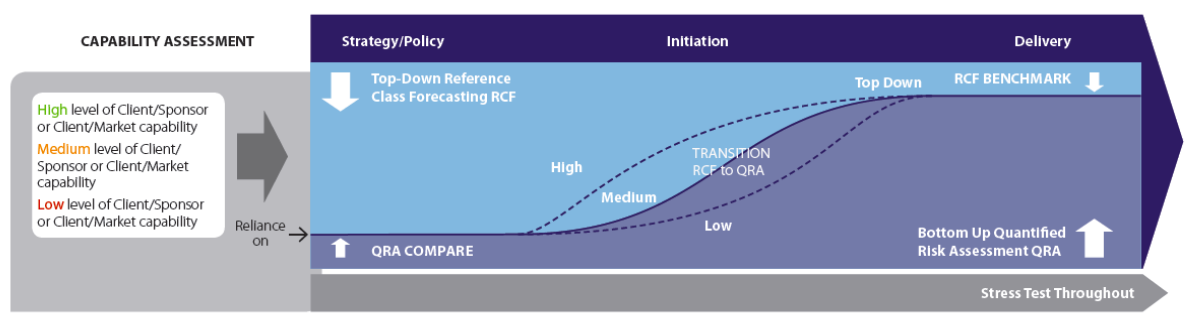

Third, project data provided need to be continuously updated to adequately reflect improvements in cost and schedule risk estimation and project delivery.

## 4.2.2  Independent review and audits of projects

The above analysis of the root causes concluded that optimism bias and political bias are both deception, but where the latter is deliberate, the former is not. While optimism can be addressed through taking the outside view, de-biasing for political bias requires additional steps.

Political bias arises when agency problems are present, i.e. when principal-agent relationships are misaligned by hidden agendas, hidden action and/or hidden information.

Incentives need to be aligned because principal and agent behavior is driven by self-interest. In order to receive accurate and un-biased forecasts, the principal (i.e. the project funder or owner) should introduce positive and negative incentives for the agent (i.e. the project forecaster) linked to the accuracy of forecasts.

In addition, hidden agendas, hidden action and hidden information pose a hazard to the principal-agent relationship and may lead to political bias.

In many megaprojects, government acts as both promoter of a project and the guardian of public interest issues for that project, such as protection of the environment, safety and of the taxpayer against unnecessary financial risks. These often conflicting objectives not only create conflicts of interest and principal-agent problems but also political bias.

Project reviews and audits can surface potential political bias in projects, e.g. the suppression of bad news. For reviews to effectively provide checks and balances, the reviews and audits need to be independent, i.e. free of political bias themselves. At a minimum, this requires reviews to be independent of any government agency overseeing a project (e.g. reviews by the national auditor), if not independent from government altogether.

## 4.2.3  Peer-review of projects

The UK Government has introduced a system of peer reviews of projects to reduce the cost and reliance on external personnel to conduct reviews.

The key advantage of these reviews is that they come from a trusted source, who has knowledge of how projects are delivered within the local government context. The reviews have sufficient independence to surface key issues and their informal, i.e. partly unminuted, nature means that thorny issues can be discussed.

In effect, the peer reviews bring an outside view to the project and thus offer an opportunity to spot signs of optimism and political biases.

# 4.3 Accountability

In addition to better oversight and independent reviews, principles of good accountability and transparency can help to address potential principal-agent problems right from the start, in order to prevent bias in estimates and forecast in the first place.

## 4.3.1 Skin-in-the-game by forecasters

In many projects forecasters are not accountable for the accuracy of forecast they produce. Incentives are often not in place to ensure that their forecasts are accurate. In most projects, incentives actually encourage inaccurate forecasts.

For example, if in a project the forecasters are paid upon successful project approval, they have strong incentives to introduce political bias into the project, which increases the chances of funding by overestimating the benefits and underestimating the cost of projects. For example, deliberately slanted forecasts have been reported in a quarter of demand forecasts for rail projects (Flyvbjerg et al. 2005).

Similarly, when project promoters want to increase the chance of funding their incentives are stacked in favor of underestimating cost and timelines in order to attract funding.

When contracts are awarded fully or with a high weighting on low price, contractors are incentivized in favor of underestimating cost and timelines in order to win contracts.

The court case between Macquarie and Syncora (Syncora Guar. Inc. v Alinda Capital Partners LLC, 2017) exposed that Macquarie paid success fees to Maunsell, the traffic forecaster in this case, for the successful issuance of bonds by Syncora. The court ruling quotes a witness describing this as unsettling behavior, yet did not award punitive damages in this case.

In similar lawsuit, AECOM, the forecasters of the bankrupt Clem Jones Tunnel in Australia, settled a claim over misrepresented forecasts for AUD 280 million (Stacey 2015).

What these cases show is that forecasters need to be held accountable for the accuracy of their forecasts. The cases also show that this should not only be a consideration after a project has failed, but accountability should be built into the forecasters' incentives and obligations right from the start of the planning process.

## 4.3.2 Accountability of decision makers and project managers

Accountability of decision makers and politicians tends to be a problem, particularly to hold those responsible for scope decisions accountable for scope creep in projects, which is a major cause of cost and schedule overruns.

In 2014, the UK Government changed the accountability structure of its public projects. Previously, if a project breached the authorized expenditure limit, the Parliament's Select Committee, which is ultimately responsible for authorizing funding, only interrogated civil servants on behalf of their ministers.

Parliament introduced a change to committee rules, so that committees can insist on hearing evidence from named officials, i.e. civil servants, even against the wishes of the responsible minister. This rule change made Senior Responsible Officer of the project, i.e. the senior project managers, directly accountable to Parliament.

# 4.4 Transparency

## 4.4.1 Reporting of project performance

During delivery, effective governance needs to provide constant challenge and control of the project. To provide adequate challenge and control, the governance bodies need to receive unbiased and up-to-date information about project performance compared to its baseline. Thus reporting enables problem solving, including quickly getting the project back on track, whenever it begins to veer off course.

Effective governance relies on multiple channels of information to senior decision makers; for example, data-driven reports on project performance and forecasts combined with reports by the management team and independent audits (Flyvbjerg and Kao 2014).

In reports, special emphasis must be placed on detecting early-warning signs that cost, schedule and benefit risks may be materialising, as they tend to do, so damage to the project can be prevented. When early-warning signs emerge, projects should revisit their assumptions and reassess cost and schedule risk and review optimism bias forecasts.

During project execution, cost and schedule variation are typically reported against the latest approved baseline. The practice aims to hold contractors accountable for the contracts they signed and hold project management accountable for their most recent plan to complete the project.

This practice ensures accountability during delivery, however, it should be extended to the planners and forecasters. Planners and forecasters should be held accountable for their forecasts by measuring and reporting the extend of underestimation or overestimation in their forecasts.

Similarly, the quality of decision making can only be revisited and improved when decision makers are held accountable for the data they based their decision on.

Measuring and reporting project performance against multiple and clearly defined baselines is needed to hold forecasters accountable for their forecasts, hold decision makers accountable for the quality of their decisions, hold project teams accountable for their project execution and contractors accountable for their contracts. Only then can enhancement and improvement of forecasting take place.

In addition, one of the largest challenges of using better forecasting methods, such as Reference Class Forecasting, is the lack of high-quality data. Thus owners and funders should make baseline and progress data internally available.

## 4.4.2 Reporting of cost effectiveness of projects

When projects are held accountable for meeting budget and timeline targets, incentives are created to increase budgets. While contingencies are always needed and unbiased analysis shows that contingencies need to be larger than typically applied, in theory, creating transparency about cost and schedule overruns might lead to projects inflating their estimates above a reasonable level.

Taking the outside view and Reference Class Forecasting, as described above, can establish what reasonable levels of contingency are.

In addition, Reference Class Forecasting could be applied to the value-for-money question, i.e. to unit cost estimates. This could provide a safeguard against unreasonably inflated cost estimates. Similarly, Reference Class Forecast applied to planned productivity could safeguard against unreasonably inflated schedule estimates.

Further, incentives schemes for unit cost and productivity targets could be based on the same logic of tiered contingency regimes, described above, when the analysis is applied to unit cost and productivity estimates. Targets set based on de-biased estimates and full distributional information better balance realism and ambition.

Thus, public sector organizations, should at the very least share internally with their planners the underlying unit cost in their projects to provide data for better inside view forecasts.

# 4.5 Other recommendations

## 4.5.1  Smart scaling

Our analysis of hydro-electric projects shows (Ansar et al. 2017) that long construction times are associated with higher cost overruns. The implication is to build as fast as possible. However, in order to go fast during construction more consideration needs to be given to this during the planning phase. Shortening the planning phase to shorten the project duration leads to project failure. Instead, projects need to go slow to go fast.

In addition to slowness, one-off projects perform worse. Case studies, e.g. the Madrid Metro extension by Manuel Melis, show that the best way to build large projects is to make them modular and deliver modules as fast as possible. Modularization within a larger program of work, like the Madrid Metro, leads to better economies of scale and, more importantly, better economies of learning.

## 4.5.2  Masterbuilder development

In many countries, the career development of project managers is hampered by the lack of recognition of the project delivery profession. Career advancement and development of civil servants tends to be tied to policy development but not delivery. In addition frequent turn-over of civil servants in project roles leads to loss of capabilities and limited opportunities for learning and development of project management specialists (Brown 2013).

Thus, the maturity of project leadership in the public sector is perceived to be lower than the leadership in the private sector partners delivering a project. This points to the need to equip project managers to effectively and efficiently lead major projects. For example, the UK government has invested in the Major Projects Leadership Academy; the Hong Kong government has a similar programme as part of the Civil Service College.

In addition, a key practical challenge is that top-management governance bodies often include representatives without prior experience in managing major projects. Thus, effective communication between the project and its governance bodies becomes a challenge. Closing this capability gap requires not only capability building on the side of the project managers but also the project sponsors.

## 4.5.3  Private finance

Public financing and financing with a sovereign guarantee are often seen as less costly and less risky than private finance, because of the lower risk premium involved in the former type of financing compared with the latter. However, public financing or financing with a sovereign guarantee does not reduce risk or costs of risk. It only transfers risk from lenders to taxpayers and is likely to increase the total risks and costs of a project.

The decision to go ahead with a project should, where at all possible, be made contingent on the willingness of private financiers to participate without a sovereign guarantee for at least one third of the total capital needs. This could result in more realistic risk assessment, a possible reduction of risk and a shift in risk from ordinary citizens to groups better able to protect themselves against risk. The pressure on performance would be higher as lenders and possible shareholders and stock market analysts would monitor the project.

The participation of risk capital does not mean that government gives up or reduces control of the project. On the contrary, it means that government can more effectively play the role it should be playing, namely as the ordinary citizen's guarantor for ensuring concerns are met about safety, environment, economics and distribution of risk.

## 4.6 Summary of the Recommendations

The data on hydro-electric dam projects and other projects, in Canada and elsewhere, show that a project's cost and schedule are frequently and systematically underestimated. The root causes of these underestimations can be found in optimism and political biases.

Planners, owners and decision makers can take steps to de-bias their projects up front and during project delivery.

One reason why unchecked optimism bias persists in many projects is that they are planned and managed with an inside view. Projects should take an outside view and compare the project to the experience of similar, past projects in order to guide planning and decision making.

In addition, to counter optimism bias, decision-makers should consider the full distributional information for relevant outcomes like cost and schedule. Decision makers should set multiple goals and targets according to their risk appetite for each goal. Thus projects can be ambitious where optimism is merited and conservative where realism is needed.

Reference Class Forecasting is a method to systematically take the outside view by using data from previous projects and thus bypassing optimism and political bias.

Additional steps are needed to correct political bias. The cause of political bias is found in principal-agent relationships, which are motivated by self-interest.

Stage-gate approvals during project planning and delivery should mandate steps to de-bias estimates; independent and peer reviews also help in surfacing signs of optimism and political biases.

Project planners and managers need to be held accountable for the plans and estimates they produce, just as structures and processes are in place to hold contractors accountable for their contracts. This should be done by introducing positive and negative incentives for planners and forecasters to produce accurate estimates. International experience shows that project accountability is often diffuse, thus the accountability placed on policy makers and planners should be extended to the project leaders delivering projects.

A further recommendation to reduce the likelihood of adverse principal-agent relationships is transparency. Enhancements in project reporting are needed, especially consistent measurement against baselines and tracking for early-warning signs that things are going wrong. This enables problem solving, including quickly getting the project back on track, whenever it begins to veer off course, which invariably happens.

However, greater internal transparency about cost and schedule overruns might create incentives to pad project budgets and timelines. Better forecasting methods, such as Reference Class Forecasting, suggest realistic contingencies. Extending the application of Reference Class Forecasting to unit cost and productivity estimates allows setting realistic targets and aligns incentives to ensure projects deliver value for money.

The data show that the best approach to deliver projects on budget is to deliver them on time, in as short a time as possible. However, to accelerate project delivery projects need to plan more carefully – shortening project durations by cutting planning short will, most likely, backfire. In addition to delivering projects fast, project approaches need to be chosen for economies of scale and learning. Smart scaling of projects, i.e. delivery at speed and with a high degree of modularization, implements both strategies and de-risks projects.

Finally, the maturity of project leadership, especially in the public sector, is often perceived to be low. Thus project leaders need to be equipped to effectively and efficiently lead projects, which often are technically, structurally or politically difficult to manage. Similarly, project sponsors and decision makers need to be equipped with the needed capabilities to understand and challenge projects in order to provide effective oversight.